# Temperature Dependence of the Cyclotron Mass in n-Type CdS

J. T. Devreese[a,*], V. M. Fomin[a,b], V. N. Gladilin[a,b], Y. Imanaka[c], and N. Miura[d]

[a]*Theoretische Fysica van de Vaste Stof, Universiteit Antwerpen (U.I.A.), Universiteitsplein 1, B-2610 Antwerpen (Belgium)*
[b]*Fizica Structurilor Multistratificate, Universitatea de Stat din Moldova, str. A. Mateevici 60, MD-2009 Chisinau (Moldova)*
[c]*National Research Institute for Metals, 1-2-1 Sengen, Tsukuba-shi, Ibaraki 305, Japan*
[d]*ISSP, University of Tokyo, Roppongi, Minato-ku, Tokyo 106, Japan*

**Abstract**

Recent cyclotron resonance experiments in n-type CdS at ultra-high magnetic fields have revealed a pronounced maximum of the electron cyclotron mass as a function of temperature. In order to interpret these data, we calculate the magneto-absorption spectra of polarons in n-CdS using the arbitrary-coupling approach. We show that in high magnetic fields the polaron effects beyond the weak-coupling approximation clearly reveal themselves in the magneto-optical absorption even at relatively small values of the Fröhlich coupling constant. In particular, those effects result in a non-monotonous behaviour of the cyclotron mass as a function of temperature. We extend the theory to take into account a combined effect of several scattering mechanisms on the magneto-absorption spectra. The extended theory allows us to interpret quantitatively the experimentally observed behaviour of the cyclotron mass in CdS.

## 1. Introduction

Measuring the magneto-absorption spectra gives a powerful tool to study the polaron effects in semiconductors. In particular, the cyclotron mass of electrons in II-VI compounds is strongly influenced by the electron-phonon interaction. Earlier experiments on II-VI semiconductors were performed at photon energies lower than the LO phonon energy (see, for example, Refs. [1,2]). Recent progress in high magnetic field technology has provided the possibility to study cyclotron resonance over a wide range of photon energies and temperatures. In the experiments of Miura and his co-workers on cyclotron resonance in bulk n-type CdS [3] and ZnCdSe/ZnSe multi-quantum wells [4] at ultra-high magnetic fields (up to 450 T), a pronounced maximum of the cyclotron mass $m^*$ as a function of temperature $T$ has been observed. With lowering $T$ from the room temperature, the cyclotron mass first increases, while below a certain temperature (about 80 – 100 K in the experiment [3]), a sudden decrease of $m^*$ occurs.

In order to interpret the experimental data [3], an interplay between several scattering mechanisms was suggested in Ref. [5], where the weak-coupling approach was used. Based on the all-coupling theory of Devreese and his associates (cf. [6]), here we show that the polaron effects beyond the weak-coupling approximation significantly influence the magneto-absorption at high magnetic fields (even for the Fröhlich coupling constant *a* as small as 0.527 in n-type CdS [7]).

---
[*] e-mail: devreese@uia.ua.ac.be

## 2. Model

Following the approach developed in Ref. [8], the magneto-optical absorption spectrum in the Faraday (active-mode) configuration is given by

$$\mathrm{Re}\,s(w,w_c,T) \propto -\frac{\mathrm{Im}\,\Sigma_+}{[w - w_c - \mathrm{Re}\,\Sigma_+]^2 + [\mathrm{Im}\,\Sigma_+]^2}, \quad (1)$$

where *s* is the electric conductivity, *w* is the frequency of the absorbed radiation, $w_c = eB/m_b c$ is the cyclotron frequency, *B* is the magnetic field, $m_b$ is the band mass of the conduction electron, and $\Sigma_+$ is the memory function defined in [8].

In Ref. [8], the memory function $\Sigma_+$ of an electron, which interacts with LO phonons, was derived using the Feynman polaron model, generalised in Ref. [9] to the case when a static magnetic field is present.

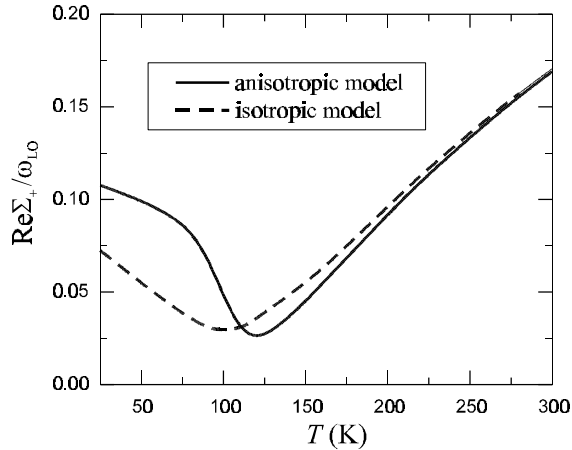

Fig. 1. Real part of the memory function for a polaron in n-type CdS (in units of the LO-phonon frequency $w_{LO}$) as a function of temperature at $w/w_{LO} = 1.9$ and $w_c/w_{LO} = 1.5$. The results are shown for two cases when the anisotropic or isotropic versions of the model [9] are used.

Within the model [9], the trial action of Feynman's polaron theory is characterised by four parameters ($v_\parallel$, $w_\parallel$, $v_\perp$, and $w_\perp$) what allows one to take into account the anisotropy of the effective electron-phonon interaction in a magnetic field. The variational parameters $v_\parallel$, $w_\parallel$, $v_\perp$, and $w_\perp$ are determined by minimising the free energy of a polaron. The theory developed in Ref. [8] can be applied to calculate the magneto-absorption spectra of polarons at arbitrary electron-phonon coupling constant, temperature, and magnetic field.

The effect of the electron-phonon interaction on the cyclotron mass can be qualitatively understood by analysing the behaviour of the function $\mathrm{Re}\,\Sigma_+$. In the temperature range from 80 to 120 K, the function $\mathrm{Re}\,\Sigma_+$, calculated for a polaron in n-type CdS, rapidly increases with lowering temperature (solid curve in Fig. 1). This increase correlates with a sudden change in the parameters $v_\perp$ and $w_\perp$, which are relevant to the polaron degrees of freedom in the plane normal to the magnetic field, and reflects a phase transition predicted in [9] for polarons in strong magnetic fields. The existence of a minimum in $\mathrm{Re}\,\Sigma_+(T)$ is a qualitatively new result of the arbitrary-coupling theory [8] as compared to the weak-coupling approximation of Ref. [5]. Note that such a minimum appears also when using the isotropic version of the model [9] with $v_\parallel = v_\perp$ and $w_\parallel = w_\perp$ (dashed curve in Fig. 1). A minimum in $\mathrm{Re}\,\Sigma_+(T)$ should give rise to a maximum of the cyclotron mass $m^*$ as a function of temperature.

## 3. Results and discussion

As seen from the magneto-absorption spectra of a polaron, which are calculated on the basis of the arbitrary-coupling theory [8] and shown in Fig. 2, the position of the cyclotron-resonance peak is a non-monotonous function of temperature. With increasing temperature, the peak first moves towards stronger magnetic fields, while at higher temperatures a shift towards weaker fields appears. For the measurement scheme used in Ref. [3], where the photon frequency *w* is fixed while the magnetic field varies, the cyclotron mass $m^*$ is determined by the relation

$$\frac{m^*}{m_b} = \frac{\overline{w}_c}{w}, \quad (2)$$

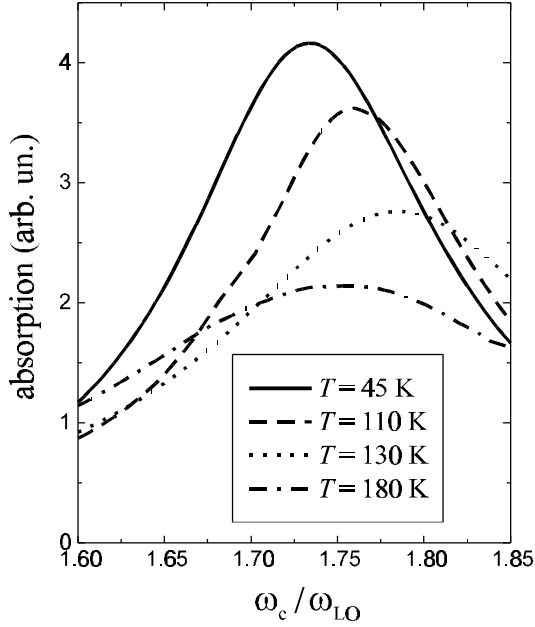

Fig. 2. Calculated magneto-absorption spectra of polarons in n-CdS as a function of cyclotron frequency at the photon wavelength $l = 16.9$ μm ($w/w_{LO} = 1.92$) and various temperatures. Only the interaction of an electron with LO phonons is taken into account.

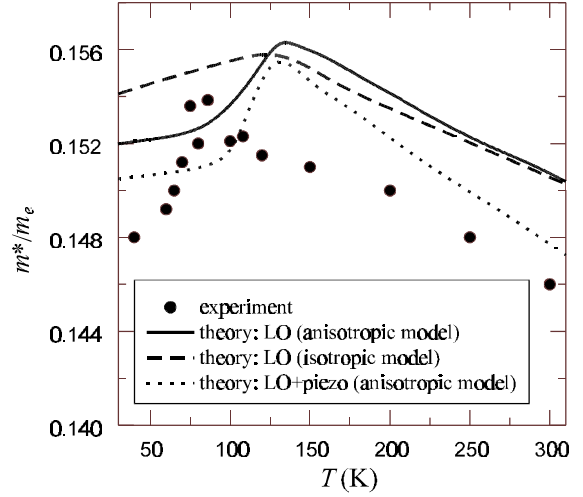

Fig. 3. Cyclotron mass, obtained from the calculated magneto-absorption spectra of polarons in n-type CdS, as a function of temperature at $l=16.9$ μm ($w/w_{LO} = 1.92$). The value $m_b = 0.168m_e$ [3] is used in the calculations. The experimental data [3] are represented by dots.

where $\bar{w}_c$ is the cyclotron frequency, which corresponds to the cyclotron-resonance peak in the magneto-optical absorption spectrum.

In Fig. 3, the solid and dashed curves show the cyclotron mass $m^*$ calculated from Eq. (2) for an electron, which interacts only with LO phonons. As distinct from the weak-coupling theory that predicts for such an electron a monotonous decrease of the cyclotron mass with increasing temperature (at $w > w_{LO}$), a more accurate approach based on the theory [8] gives for $m^*(T)$ a behaviour similar to that observed in the experiment [3]. In particular, below $T \approx 130$ K both the curves obtained using either anisotropic or isotropic polaron model [9] demonstrate a pronounced decrease of $m^*$ with decreasing temperature. However, the calculated cyclotron mass appears to be larger than the experimental values of $m^*$ over the major part of the temperature region under consideration. Furthermore, the maximum of the calculated $m^*(T)$ is shifted towards higher temperatures with respect to the experimentally observed maximum.

As the results of Ref. [5] imply, the aforementioned discrepancies between the experimental and theoretical data can be attributed to the presence of other scattering mechanisms in the experimental sample. Here we have extended the arbitrary-coupling approach [8,9] to account for a combined effect of several scattering mechanisms on the magneto-absorption spectra of polarons.

The piezoelectric interaction of electrons with acoustic phonons is known to appreciably affect the cyclotron mass in CdS [1]. The corresponding contributions to the free energy and to the memory function $\Sigma_+$ are calculated within the elastic approximation assuming that $k_B T \gg \hbar w_{piezo}$ for typical frequencies of piezoacoustic vibrations $w_{piezo}$. The value $K^2 = 0.0038$ is used for the piezoelectric-interaction constant [10]. As seen from Fig. 3 (dotted curve), this type of interaction leads to a considerable decrease of the cyclotron mass, especially at high temperatures.

Since the sample used in the experiments [3] was not intentionally doped, a relatively high concentration of electrons in this sample (3 $10^{15}$ cm$^{-3}$ at the room temperature) is supposed to be due to vacancies. The concentration of vacancies, $N$, is estimated to be $10^{16}$–$10^{17}$ cm$^{-3}$. To treat the elastic scattering of electrons by the static lattice defects, we use a gaussian potential $V_0 \exp[-k_0^2(\mathbf{r}-\mathbf{R}_i)^2]$, where $\mathbf{r}$ and $\mathbf{R}_i$ define the positions of the electron and of the $i$-th defect, respectively. The calculations are performed within the random-phase approximation.

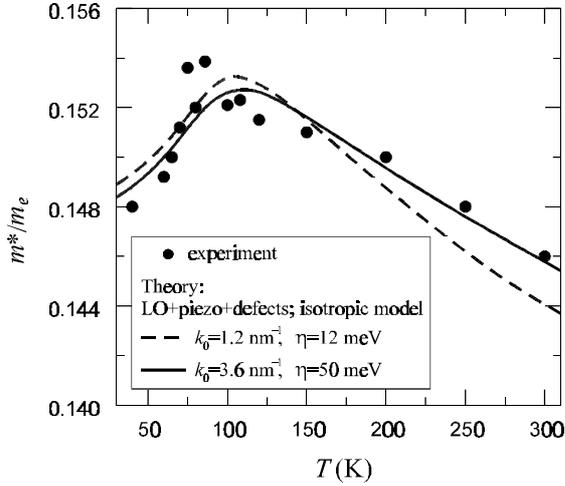

Fig. 4. Calculated cyclotron mass in n-type CdS in comparison with the experimental data [3]. Theoretical results are displayed for two sets of the parameters $k_0$ and $h \equiv V_0 \sqrt{N k_0^{-3}}$ .

The theoretical results for the cyclotron mass, which are obtained taking into account both the polaron effect and the elastic scattering of electrons by piezoacoustic phonons and lattice defects, are displayed in Fig. 4. As seen from Fig. 4, our model based on the arbitrary-coupling theory [8] can provide a quantitative interpretation for the experimentally observed behaviour of the cyclotron mass in CdS.

## 4. Conclusions

Using the arbitrary-coupling approach [8,9], we have analysed the magneto-absorption spectra of polarons in n-CdS at high magnetic fields ($w_c > w_{LO}$) as a function of temperature. Even at relatively small values of the Fröhlich coupling constant ($a$~0.5), the weak-coupling approximation occurs to be insufficient when calculating the magneto-absorption spectra of polarons in strong magnetic fields. As distinct from the weak-coupling theory, the approach [8,9] gives for the cyclotron mass of polarons a behaviour, which is similar to that observed in the experiment [3]. The approach [8,9] is extended in order to take into account a combined effect of several scattering mechanisms on the magneto-absorption spectra. The calculated cyclotron mass is in fair agreement with the experimental data for n-CdS.

## Acknowledgements

This work has been supported by the I.U.A.P., GOA BOF UA 2000, F.W.O.-V. projects Nos. G.0287.95, 9.0193.97 and the W.O.G. WO.025.99N (Belgium).

## References

[1] K. Sawamoto, J. Phys. Soc. Japan **18** (1963) 1224.
[2] K. Nagasaka, Phys. Rev. B **15** (1977) 2273.
[3] Y. Imanaka, N. Miura, and H. Nojiri, Physica B **246-247** (1998) 328.
[4] Y. Imanaka and N. Miura, in: 12th Int. Conf. *Electronic properties of two-dimensional systems* (Tokyo, 1997), Conf. Workbook, p. 467.
[5] J. T. Devreese, V. M. Fomin, V. N. Gladilin, Y. Imanaka, and N. Miura, J. Crystal Growth **214-215** (2000) 465.
[6] J. T. Devreese, in: Encyclopedia of Applied Physics, Vol 14 (VHC Publishers, Inc., 1996) pp. 383 – 413.
[7] D. L. Rode, Phys. Rev. B **2** (1970) 4036.
[8] F. M. Peeters and J. T. Devreese, Phys. Rev. B **34** (1986) 7246.
[9] F. M. Peeters and J. T. Devreese, Phys. Rev. B **25** (1982) 7281; Phys. Rev. B **25** (1982) 7302.
[10] A. A. Klyukanov and E. P. Pokatilov, Zh. Exper. Teor. Fiz. **60** (1971) 1878 [Sov. Phys. JETP **33** (1971) 10159].